\documentclass[12pt]{article}
\usepackage{epsfig}

\begin{document}

\begin{flushright}
  DESY--01--126 \\
  hep-ph/0109040
\end{flushright}

\vspace{\baselineskip}

\begin{center}
\textbf{\LARGE Elastic meson production and Compton
scattering\,\footnote{Talk given at the Ringberg Workshop on New
Trends in HERA Physics 2001, 17--22 June 2001, Ringberg Castle,
Tegernsee, Germany. To appear in the Proceedings.}  } \\
\vspace{2\baselineskip}
{\large M. Diehl}
\\
\vspace{\baselineskip}
Deutsches Elektronen-Synchroton DESY, 22603 Hamburg, Germany \\
Email: markus.diehl@desy.de
\\ 
\vspace{3.5\baselineskip}
\textbf{Abstract}\\
\vspace{1\baselineskip}
\parbox{0.9\textwidth}{}I discuss recent progress in the theory of
exclusive meson production and Compton scattering, focusing on
hard-scattering factorization and on the dipole formalism.
\end{center}
\vspace{0.5\baselineskip}
%

%==============================================================
\section{Introduction}

In this talk I will discuss selected issues in elastic meson
production and Compton scattering.  Given limited time and space I
cannot review all activity in the field since the last Ringberg
meeting two years ago \cite{teu99}; in particular I will say little
about heavy-meson production and nothing about odderon physics.  My
first topic is the dipole description of diffractive processes and its
relation with hard-scattering factorization.  More on these topics can
be found in \cite{bar01,col01,gol01}.  A second part is devoted
to scaling and its interplay with the helicity structure of
hard-scattering processes, a topic where there is now a variety of
data.  In a third part, I focus on deeply virtual Compton scattering,
i.e., the process $\gamma^*p\to \gamma p$ at large $Q^2$ and small
$t$.  I will give short summaries at the end of each part.

%==============================================================
\section{Two pictures}

\begin{figure}
\begin{center}
\leavevmode
\epsfxsize=0.45\textwidth  
\epsffile{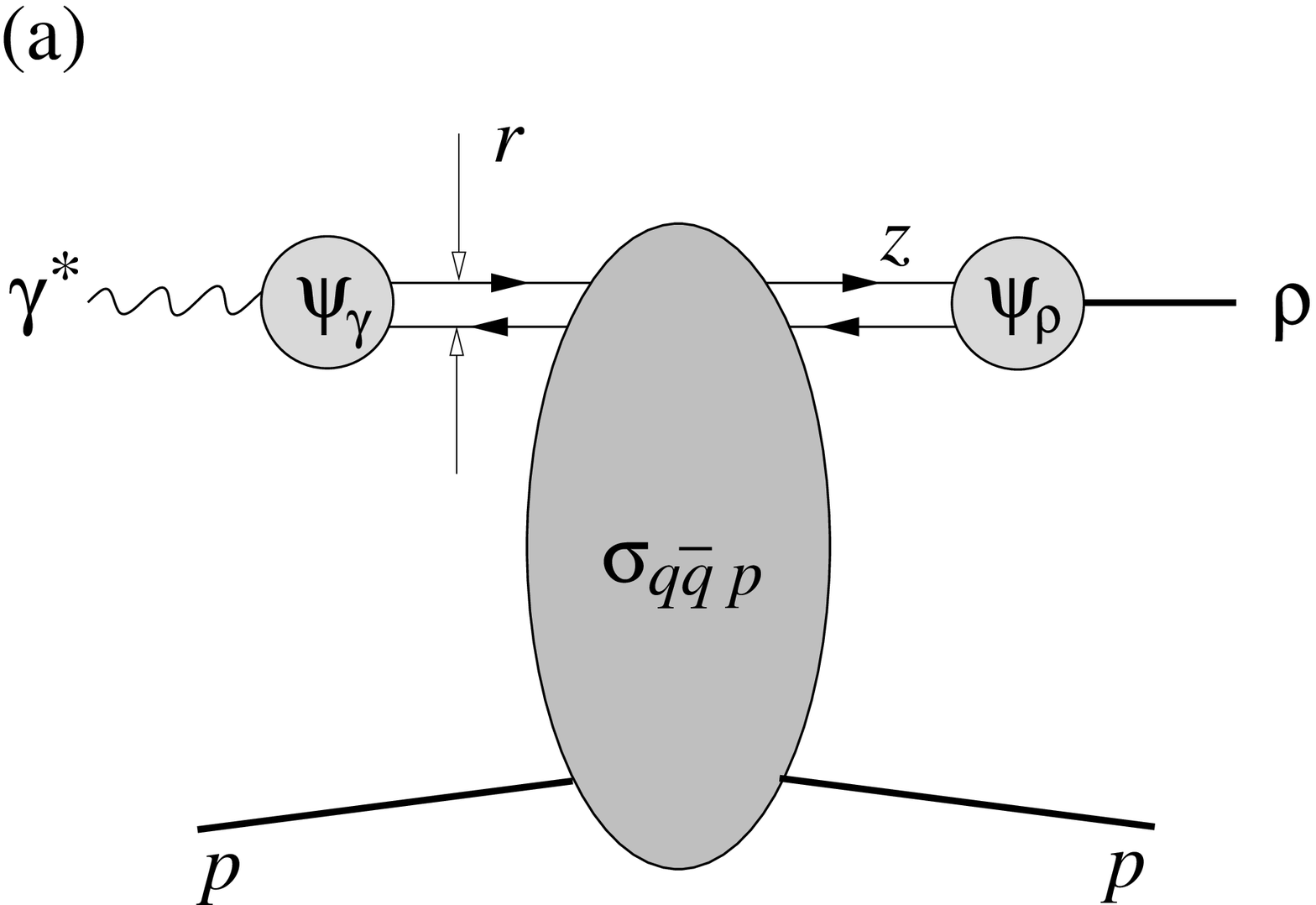}
\hspace{0.05\textwidth}
\epsfxsize=0.40\textwidth  
\epsffile{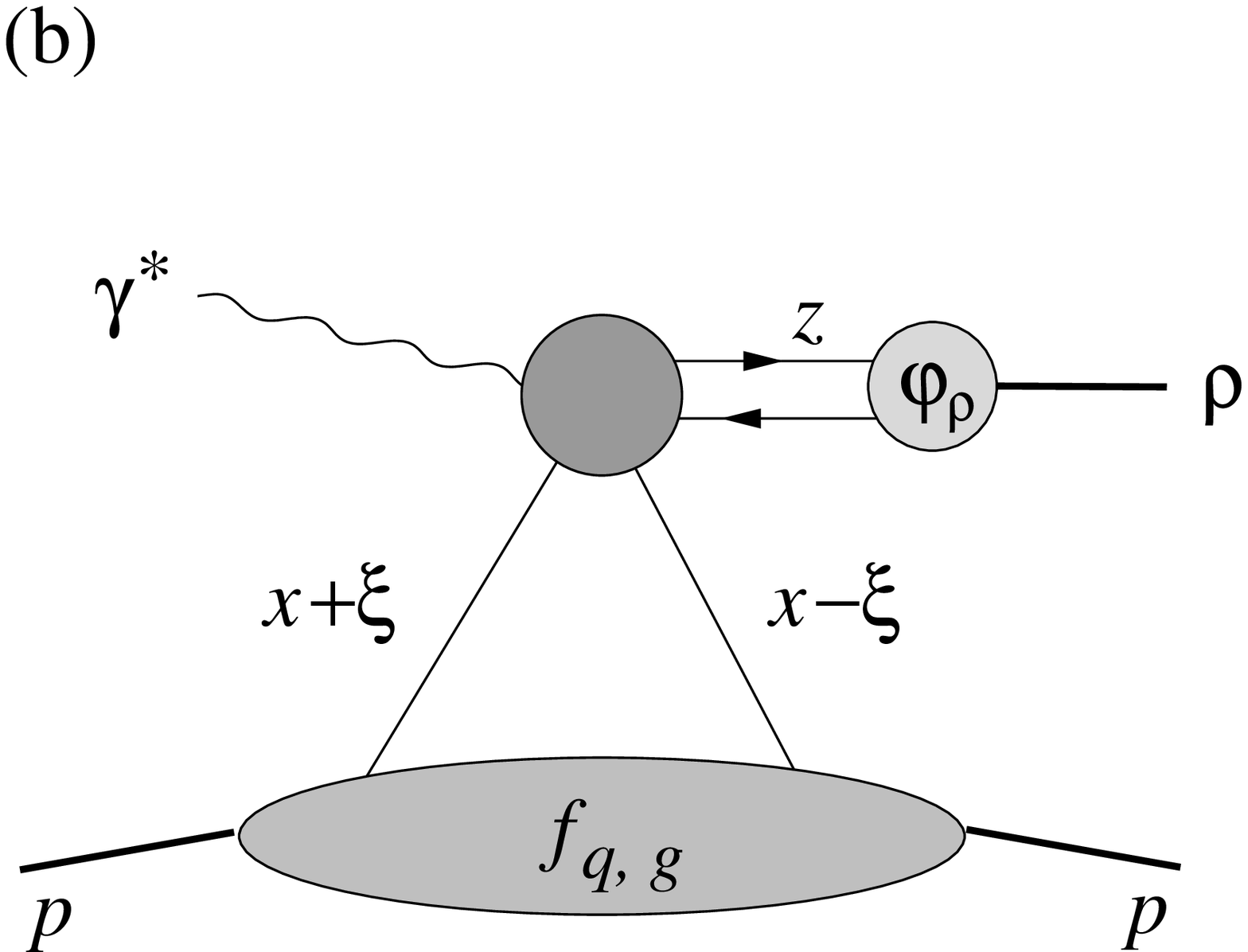}
\end{center}
\caption{The process $\gamma^* p\to \rho p$ in the dipole picture (a),
and in hard-scattering factorization (b).}
\label{fig:pic}
\end{figure}

There are two rather different ways of viewing a process like
$\gamma^* p\to \rho p$ or $\gamma^* p\to \gamma p$. Each comes with
its own formalism, region of validity, and its way of organizing the
scattering amplitude into different building blocks.  The dipole
picture is a high-energy approximation. Its ingredients, as shown in
Fig.~\ref{fig:pic}a, are light-cone wave functions $\psi_{\gamma}$,
$\psi_{\rho}$ giving the amplitude for the virtual photon or meson to
fluctuate into a $q\bar{q}$-pair, and the amplitude ${\cal
A}_{q\bar{q}\, p \to q\bar{q}\, p}$ of this $q\bar{q}$-pair to scatter
elastically off the proton.  In the high-energy approximation one
often retains only the imaginary part of ${\cal A}$ and replaces it
with the total $q\bar{q}\, p$ cross section $\sigma_{q\bar{q}\, p}$
via the optical theorem.  Hard-scattering factorization, shown in
Fig.~\ref{fig:pic}b, is valid in the limit of large photon virtuality
$Q$.  Its building blocks are a kernel describing the hard scattering
between quarks, gluons, and photons, the $q\bar{q}$ distribution
amplitude $\varphi_\rho$ of the meson, and a generalized quark or
gluon distribution $f_{q,g}$ in the proton.

\subsection{Variables and their roles}
\label{sec:pictures}

Let us look in some detail at the variables appearing in the
respective building blocks of the two formalisms.  The dipole cross
section depends on one variable related with the scattering
energy. There is an ongoing debate on whether the appropriate choice
for this is Bjorken's scaling variable $x_B$ \cite{gol99} or rather
the $\gamma^* p$ c.m.\ energy $W^2$ \cite{for99}.  $\sigma_{q\bar{q}\,
p}$~further depends on the transverse size $r$ of the $q\bar{q}$
dipole.  Note that $r$ is conserved by the interaction in the
high-energy limit, unlike its Fourier conjugate, the relative
transverse momentum $k_T$ in the $q\bar{q}$ pair.

Generalized parton distributions \cite{mul98} depend on two
longitudinal momentum fractions, which can be parameterized
symmetrically as shown in Fig.~\ref{fig:pic}b.  Here $\xi= x_B
/(2-x_B)$ is determined by the external kinematics, whereas $x$ is a
loop variable, which needs to be integrated over and thus has no
analog in the dipole cross section.  Note that generalized parton
distributions do not depend on the transverse momentum $k_T$ of the
quarks or gluons in the proton.  This is because in hard-scattering
factorization one makes a collinear approximation, i.e., one sets
$k_T=0$ in the hard-scattering kernel.  The loop integral over $k_T$
can then be performed for the lower blob in Fig.~\ref{fig:pic}b alone
and gives the parton distribution.  The upper limit of this integral
is taken of order $Q$ and provides the factorization scale of
$f_{q,g}$.

In the same fashion one makes the collinear approximation for the
relative transverse momentum $k_T$ in the $q\bar{q}$ pair.  The
corresponding loop integral is then performed for the meson wave
function alone and gives the distribution amplitude $\varphi_\rho$ at
a factorization scale of order $Q$.  Expressed in transverse position
space, this procedure means that one is sensitive only to the meson
wave function for $q\bar{q}$ separations $r$ up to order $1/Q$.

Both the dipole cross section and the generalized parton distributions
depend on the invariant momentum transfer $t$, which in both pictures
may be replaced with the impact parameter $b$ after a Fourier
transform in transverse space.  The simplest type of ansatz for the
$t$ dependence often made in models is that it factorizes, e.g.,
$\sigma(x_B,r,t) = \sigma(x_B,r)_{t=0}\; e^{B t/2}$ or $f(x,\xi,t) =
g(x,\xi)\, F(t)$ with a constant slope parameter $B$ or a general form
factor $F(t)$.  While this is a convenient first approximation it
misses important physics.  At small $x_B$ the steepening of the $t$
dependence with energy, i.e., shrinkage, has recently been advocated
as deeply connected with confinement \cite{bar00}.  In generalized
parton distributions, this phenomenon is reflected by a correlation
between the dependence on $t$ and the longitudinal momentum variables
$x$ and $\xi$.  An intuitive picture can be gained when going to
impact parameter $b$, which describes the transverse position of the
struck parton in the proton \cite{bur00}.  Representing the parton
distribution as the overlap of light-cone wave functions for the
initial and final hadron \cite{bro00}, shrinkage is then seen as a
nontrivial dependence of the transverse distribution for partons with
different longitudinal momentum in the target.  Such a picture is not
restricted to small $x_B$, and several models for generalized parton
distributions at larger $x$ and $\xi$ indicate that a factorizing
$t$-dependence is too simple \cite{pen00,die98}, which reflects a
nontrivial interplay between transverse and longitudinal degrees of
freedom in the hadron.

\subsection{Regions of validity}

The hard-scattering formalism is valid in the limit where the scale
$Q$ is large, it gives the amplitude up to corrections in powers of
$1/Q$ times logarithms of $Q^2$.  This can be proven using general
methods as explained in \cite{col01,col96}.  The hard scale $Q$ may be
replaced by a large internal quark mass, for instance in
photoproduction $\gamma p\to J/\Psi\, p$, although no formal proofs of
factorization exist for this case.  Hard-scattering factorization can
be applied for $x_B$ small or not, and thus provides a link between
high-energy phenomena at H1 and ZEUS and physics at HERMES energies.
The dipole formalism, on the other hand, requires $x_B$ to be small,
but goes beyond the leading power in $1/Q$.  Again, one may replace
the scale $Q$ from a highly virtual photon with the large quark mass
in quarkonium production.

In the joint limit of large $Q$ and small $x_B$, more precisely to
leading power $1/Q$ and to double leading logarithm, $\log Q^2$ and
$\log(1/x_B)$, both pictures claim validity and their predictions
coincide.  In this approximation the dipole cross section at small
separation~$r$ is proportional to the gluon density,
$\sigma_{q\bar{q}\, p} \sim r^2 \, \overline{x}
g(\overline{x};\overline{Q})$, where $\overline{x}$ is of order $x_B$
and the factorization scale $\overline{Q}$ of order $Q$ or $1/r$.  In
a leading logarithmic approximation we cannot make the statement ``of
order'' more precise.  To the same accuracy we can further not
distinguish between the usual gluon density and the generalized one
$g(\overline{x},\xi;\overline{Q})$ at some $\overline{x} \sim \xi$.
Numerically however, the difference between the gluon density, say, at
$Q$ and at $Q/2$, or at $x_B$ and at $x_B/2$ is quite important in the
low-$x_B$ regime, and many attempts have been made to improve on the
simple relation $\sigma^{q\bar{q}\, p} \sim x_B g(x_B,Q)$, making more
refined choices for $\overline{Q}$ and $\overline{x}$, and
incorporating the generalized gluon distribution.  This is somewhat
reminiscent of choosing the scale of $\alpha_s$ in a fixed-order
perturbative calculation: while it makes sense from a phenomenological
point of view, one should not forget that one is asking a question to
which strictly speaking there is no answer at the established accuracy
of the calculation.

In contrast to hard-scattering factorization, it is not clear for the
dipole formalism up to which parametric accuracy it is valid.  It does
emerge as a description in the leading $\log(1/x_B)$ approximation,
and has been studied in connection with the BFKL pomeron
\cite{nik94,bia00}.  Whether the dipole picture can provide a
systematic description of the scattering amplitude beyond the leading
$\log(1/x_B)$ is not known and subject of ongoing studies in the
connection with the next-to-leading logarithmic corrections to BKFL
\cite{bar01}.  If the picture persists at that accuracy, it will most
probably need to be extended in several aspects.  Beyond leading
$\log(1/x_B)$ the dipole cross section can depend on the longitudinal
momentum fraction $z$ of the quark in the $q\bar{q}$ dipole.  In
general, the transverse dipole size will not be conserved any more
\cite{bia00}.  Also, more complicated objects than just a $q\bar{q}$
dipole such as $q\bar{q}g$ states will have to be considered as
scattering off the target \cite{bar01,bar01a}.

Given the simplicity and versatility of the dipole formalism in the
form of Fig.~\ref{fig:pic}a, it is tempting to extend it beyond the
strict limits where it has been derived so far.  Incorporating the
generalized gluon distribution mentioned above goes beyond the leading
$\log(1/x_B)$, as does the inclusion of the real part of the dipole
scattering amplitude ${\cal A}_{q\bar{q}\, p \to q\bar{q}\, p}$, which
is subleading in $\log(1/x_B)$.  A perhaps even bolder extension is to
try and describe $\sigma_{q\bar{q}\, p}$ at large dipole size $r$,
where the connection between dipole cross section and gluon density
becomes unclear, and where perturbation theory ceases to be valid.  In
the same direction goes the extension in $Q$ down to the
photoproduction limit $Q=0$.  The wave function of a real photon is
nonperturbative so that at present one has to model it, as one does
for meson wave functions.  One should however be aware that for a
purely soft transition such as $\gamma \to \rho$ the contribution from
higher Fock states in the photon is no longer suppressed by a small
$\alpha_s$ compared with the $q\bar{q}$ component, and that a Fock
state expansion on current quarks and gluons may not be practical at
all.  At this point one must invoke further arguments, e.g.\ along the
lines of \cite{nac91} or of \cite{rad95}.

\subsection{Applications}

There is one particularly interesting phenomenon that can readily be
incorporated into the dipole formalism while clearly going beyond the
leading-twist physics of hard-scattering factorization.  This is the
particular correlation of the $r$ and $x_B$ dependence of
$\sigma_{q\bar{q}\, p}$ termed saturation, discussed in more detail in
\cite{gol01}.

One of the strategies pursued is to take a particular model of the
dipole cross section, compare its results to data, and see if one is
sensitive to saturation effects in $\sigma_{q\bar{q}\, p}$.  The
dipole model has been applied to light and heavy vector meson
production in~\cite{cal01}, and $J/\Psi$ photoproduction was studied
in \cite{got01} and \cite{fra00}.  It is interesting to note that the
last two studies both achieve agreement with the data, but with
different physics mechanisms.  Whereas the main ingredient ensuring a
good description in \cite{got01} were shadowing corrections, the
dominant effect in \cite{fra00} came out to be the choice of
factorization scale in the gluon density when expressing the dipole
cross section at small~$r$.  Given our incomplete ability to reliably
calculate either of these effects, I conclude that with present theory
and with the kinematical range of present data the relevance of
saturation cannot be firmly established.

The reverse approach has been followed in \cite{mun01}, where an
attempt was made to reconstruct the dipole cross section from $\rho$
production data.  Unfortunately, it was found that with the $t$-range
of the data no reliable extraction of $\sigma_{q\bar{q}\, p}$ was
possible for small impact parameters $b$, where saturation effects can
be expected to set in first.

In conclusion, hard-scattering factorization and the dipole formalism
are valid in different kinematical regimes, but these regimes overlap
and there the two pictures give complementary descriptions.  The
dipole picture provides a versatile framework for describing
diffractive reactions, and many studies are trying to refine it with
quite good phenomenological success.  What we are however lacking at
this time is a theoretical understanding of what the limits of
applicability are for this formalism and whether such refinements can
be put on a systematic basis.

%==============================================================
\section{Scaling and helicity}

\subsection{Scaling}

I focus now on the hard-scattering formalism.  To remind us, the
factorization theorem, e.g.\ for $\rho$ production states that the
amplitude for $\gamma^*_L\, p\to \rho_L^{\phantom{*}}\, p$ scales like
$1/Q$ times logarithmic corrections in $Q^2$, in the limit of large
$Q^2$ and at \emph{fixed} $t$ and $x_B$.  The latter point is
important to remember when looking at plots of the $Q^2$ dependence
integrated over a fixed interval not in $x_B$ but in the scattering
energy $W$.  While this is rather natural from the point of view of
the experimentally accessible phase space, it does not allow one to
directly check the predicted scaling laws, except for quantities whose
dependence on $W$ is rather flat.  Notice also that the logarithmic
corrections to the predicted power behavior can be large at small
$x_B$. They certainly are for the inclusive structure function
$F_2$---one would not have discovered Bjorken scaling from the HERA
small-$x_B$ data.  A fit to preliminary data on the cross section
$\sigma_L$ for longitudinal $\rho$ production, giving a $Q^2$ behavior
like $(Q^2 + m_\rho^2)^{-n}$ with $n=1.89\pm 0.03$ for $Q^2>2
\mbox{~GeV}^2$ \cite{cri01} and $n=2.21\pm 0.03$ for $Q^2>5
\mbox{~GeV}^2$ \cite{zeu00}, is by itself not in contradiction to the
scaling prediction that $\sigma_L$ should go like $1/Q^6$ times
logarithmic terms.

On the theory side we are far from a systematic understanding of the
$1/Q^2$ power corrections in the leading amplitudes such as
$\gamma^*_L\, p\to \rho_L^{\phantom{*}}\, p$, $\gamma^*_L\, p\to \pi
p$, $\gamma^*_T\, p\to \gamma p$.  There are however estimates based
on particular mechanisms. Recently the effects of finite parton $k_T$
in the hard scattering part of these processes were estimated in
\cite{van99} and found to be rather substantial in the case of meson
production.  For DVCS, the power corrections due to the hadronic
component of the outgoing real photon were estimated in \cite{don00}
using vector dominance, giving contributions of order 10\% to 20\% at
amplitude level for the kinematics of the H1 and ZEUS data
\cite{lob01}.

\subsection{Hadron helicity selection}

Factorization theorems only tell us how to calculate selected helicity
amplitudes for a process, but they do include the statement that all
others are suppressed by powers of $1/Q$.  Let us review the general
ingredients that go into this helicity selection at large scales.  The
first is the collinear approximation for the $q\bar{q}$ pair forming a
meson, described in section~\ref{sec:pictures}.  In this approximation
the quark and antiquark leaving the hard scattering process are
strictly collinear, so that the spin along their direction of motion
is only due to their helicities. Because of angular momentum
conservation, the helicity of the produced meson must be the same,
i.e., the sum of the $q$ and $\bar{q}$ helicities.  This no longer
holds at the level of $1/Q$ suppressed terms, where one takes $k_T$ in
the hard scattering into account: then the $q\bar{q}$ pair can carry
orbital angular momentum.  Note that at the same level of accuracy one
must also consider the $q\bar{q}g$ Fock state of the meson, where the
gluon carries helicity in addition to the quarks.  In transverse
position space the collinear approximation means that one neglects
$1/Q$ compared with the characteristic distance over which the meson
wave function varies significantly.  For a photon with its pointlike
component, one cannot make this approximation.  In fact, photons are
directly attached to the hard scattering subprocess and one does not
set to zero the relative $k_T$ of the $q\bar{q}$ pair they couple to.
The hadronic component of the photon, which may contribute at the
level of power corrections, can however be treated in a similar way as
a meson, with corresponding light-cone wave functions and distribution
amplitudes.

The second ingredient for helicity selection is that in a
hard-scattering subprocess the chirality along a light quark line is
conserved, since the light quark masses are neglected to the accuracy
in question.  Notice that this is not applicable for heavy quarks,
unless $Q$ is much larger than their mass.  Chirality breaking due to
the axial anomaly does not spoil this type of argument since it does
not affect hard-scattering kernels \cite{col99}.

\begin{figure}
\begin{center}
\leavevmode
\epsfxsize=0.5\textwidth  
\epsffile{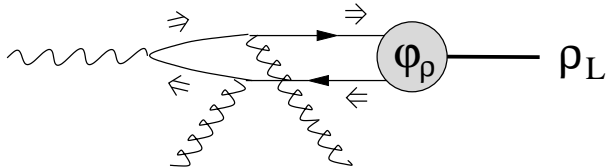}
\end{center}
\caption{Formation of a longitudinal $\rho$ from the hard-scattering
process $\gamma^* g\to q\bar{q}\, g$.}
\label{fig:rho}
\end{figure}

Consider now $\rho$ electroproduction at small $x_B$, which is
dominated by the hard subprocess $\gamma^* g\to q\bar{q}\, g$.  As
shown in Fig.~\ref{fig:rho} the helicities of the $q\bar{q}$ pair
forming the $\rho$ must be antialigned, so that longitudinal $\rho$
production is leading at large $Q^2$.  To make a transverse $\rho$
costs a factor of $1/Q$ in the amplitude.  Scaling thus predicts that
the ratio $R=\sigma_L /\sigma_T$ of longitudinal and transverse cross
sections should be linear in $Q^2$ at large $Q^2$, but again only up
to logarithmic effects.  If we strictly demand $R\propto Q^2$ we are
further assuming that these effects, where they are strong in the
cross section, should be similar for both polarizations.  This need
not be the case, and the calculation for small $x_B$ of Martin et al.\
\cite{mar99} for instance finds them clearly different, although not
much.  Data at moderate $x_B$ are interesting in this context, because
there one would not expect large logarithmic effects in
$Q^2$. Unfortunately, the existing data does not cover a large span in
$Q^2$ for values above which one may think of applying hard-scattering
arguments.  Given the still large statistical errors in the
high-energy data \cite{cri01,zeu00} I find it difficult to see whether
$R$ is approximately linear in $Q^2$ for values above, say, 2 to
3~GeV$^2$.

The selection of longitudinal $\rho$ polarization should also hold in
the subprocess $\gamma g \to q\bar{q}\, g$ at large momentum transfer
$t$ between the real photon and the $q\bar{q}$ pair, which is believed
to dominate the process $\gamma p\to \rho\, Y$ where the proton
dissociates into a hadronic system $Y$.  The preliminary results from
ZEUS \cite{cri01,zeu01} are thus puzzling, since transverse $\rho$
clearly dominate even up to $|t|$ of several GeV$^2$.  Ivanov et al.\
\cite{iva00} have suggested that the reason might be a numerically
large power correction due to the hadronic part of the photon.  This
part has a wave function component where the $q$ and $\bar{q}$ have
equal helicities, so that according to our the above arguments they
will form a transverse $\rho$ after the hard scattering.  It will be
interesting to see if this mechanism can explain the $t$-dependence
and normalization of the data at large momentum transfer.

\subsection{Photon helicity selection}

The selection of photon helicity comes about in a slightly different
way than for hadrons.  Going back to the example of $\rho$
electroproduction, we recall that in the hard-scattering kernel one
also makes a collinear approximation for the partons coming from the
proton side, and further neglects $t$ compared with $Q^2$.  It then
follows that the partons from the proton transfer no spin along the
direction of the $\gamma^*$ and $\rho$ momenta, except if they change
their helicity.  Generalized quark and gluon distributions with
helicity flip have indeed been studied, as well as the helicity
amplitudes in DVCS and meson production where they contribute
\cite{hoo98,bel00}.  Except when such distributions can contribute,
one obtains $s$-channel helicity conservation with the above
arguments, and since in $\rho$ electroproduction the meson is
longitudinal at large $Q$, the photon must be, too.  The same holds
for the production of a spin-zero meson such as a $\pi$.  At the level
of power corrections, this selection rule fails since one can have
transfer of orbital angular momentum from the transverse momentum of
the partons, or of helicity from additional partons exchanged between
the proton and the hard-scattering subprocess.

Let me add that the inclusion of parton $k_T$ in the hard scattering
does not automatically lead to a change of helicity between the photon
and meson.  In the high-energy or $k_T$-factorization scheme, for
instance, one picks up the leading $\log(1/x_B)$ terms in the
amplitude and does include gluon $k_T$ in the hard scattering, but the
contribution providing the leading $\log(1/x_B)$ does not transfer any
angular momentum.  In this case one also retains the transition
$\gamma^*_T\, p \to \rho_T^{\phantom{*}}\, p$ at high energy, although
it is subleading in the large-$Q^2$ limit.

Going back to moderate energies, the HERMES data for exclusive $\pi^+$
electroproduction \cite{tho01,bia01} on a longitudinally polarized
target shows a clear $\sin\varphi$ signal in the distribution of the
angle $\varphi$ between the lepton and hadron planes in the target
rest frame.  The effect is too large to be explained as due to the
small transverse spin of the target relative to the exchanged
$\gamma^*$, so that at least part of it must come from the
interference between amplitudes with transverse and longitudinal
polarization of the intermediate $\gamma^*$.  We can infer that both
amplitudes which are leading and nonleading in the large $Q$ limit are
visible in the kinematics of the experiment.  Furthermore, not all
possible combinations of such interference terms are large.  The
single-spin asymmetry data shows no indication of a $\cos\varphi$
modulation in the unpolarized cross section appearing in its
denominator.  Such a term would go with the \emph{real} part of an
interference term \emph{averaged} over the initial proton helicity.
The observed $\sin\varphi$ signal involves the \emph{imaginary} part
of the interference term for the \emph{difference} of the two target
helicities.  We can at present not decide whether the $\cos\varphi$
term is small because of the phase between the interfering amplitudes
of because of their dependence on the proton spin.  Future data, also
with transverse target polarization, should bring us closer to
understanding the pattern of the different helicity transitions.

\begin{figure}[b]
\begin{center}
\leavevmode
\epsfxsize=0.65\textwidth  
\epsffile{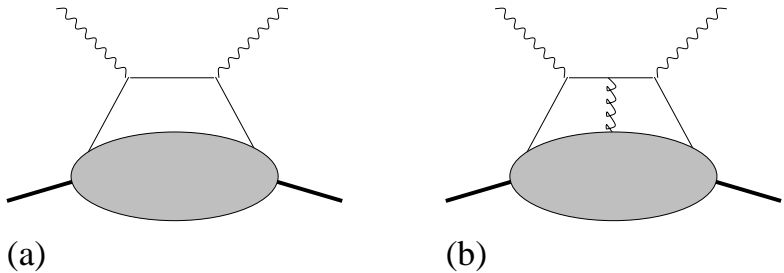}
\end{center}
\caption{a) Quark handbag diagram for DVCS. b) Diagram with an
additional exchanged gluon.}
\label{fig:compton}
\end{figure}

While there is at present no theory description of power suppressed
helicity amplitudes in pion production, the situation is better for
DVCS.  The final state photon is transverse, so that the initial one
must also be transverse to leading order in $1/Q$.  Transitions
suppressed by just one power in $1/Q$ involve a $\gamma^*_L$, and we
have now a systematic description of these to leading order in
$\alpha_s$ \cite{ani00}, closely following the analog of the spin
dependent structure function $g_2$.  Broadly speaking, they are due to
two types of contributions.  One involves the same handbag diagrams
that give the leading amplitude with a $\gamma^*_T$, see
Fig.~\ref{fig:compton}a, but with one unit of orbital angular momentum
transferred by the quarks as described above.  Using the equations of
motion for the quark fields, the corresponding proton matrix elements
can be reduced to the same generalized parton distributions that
describe the leading amplitude.  This is the contribution retained in
the so called Wandzura-Wilczek approximation.  The other one, shown in
Fig.~\ref{fig:compton}b, involves the exchange of two quark lines and
an additional gluon between the proton and the hard scattering; it
comes with corresponding twist-three parton distributions.

Information on the different photon helicity amplitudes is again
contained in the $\varphi$ dependence of the $ep\to ep\gamma$ cross
section.  At sufficiently large $Q^2$, a $\sin\varphi$ distribution of
the lepton polarization asymmetry is due to a $\gamma^*_T$ and a
$\sin(2\varphi)$ is the signature of a $\gamma^*_L$, up to at least
partly calculable corrections in $1/Q$ \cite{die97}.  The data from
HERMES \cite{bia01,air01}, and also from a yet lower energy experiment
at Jefferson Lab \cite{ste01} is consistent with a sole $\sin\varphi$,
so that we do not seem to have a signal for longitudinal $\gamma^*$
polarization in these kinematics.

\begin{figure}
\begin{center}
\leavevmode
\epsfxsize=0.6\textwidth  
\epsffile{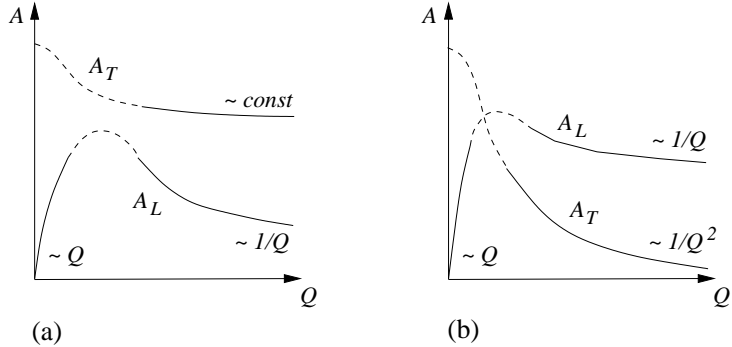}
\end{center}
\caption{Qualitative behavior of the scattering amplitudes $A_{T,L}$
for transverse and longitudinal initial photon in (a) $\gamma^*p\to
\gamma p$ and (b) $\gamma^*p\to \pi p$.}
\label{fig:pattern}
\end{figure}

Let me in this context point out a noteworthy difference between
Compton scattering and meson production.  In the photoproduction limit
$Q^2\to 0$, the transition amplitude with a $\gamma^*_L$ has to vanish
like $Q$ because of gauge invariance, so that $\gamma^*_T$ dominates
at very small $Q$.  In Compton scattering this is also the
polarization state dominating at very large $Q$, where the
$\gamma^*_T\, p$ amplitude goes to a constant and the $\gamma^*_L\, p$
one falls like $1/Q$.  The general pattern is thus the one of
Fig.~\ref{fig:pattern}a, where for simplicity I ignore the proton spin
dependence. {}From general principles we cannot say where the
longitudinal amplitude has its maximum and how large it is.  There is
no reason why it should be small compared with the transverse one
there, it might even be bigger and cross the curve for transverse
photons twice.  This depends on how fast the $\gamma^*_L\, p$
amplitude rises at small $Q$ and at which $Q$ it starts falling off
again.  For meson production, say for $\gamma^* p\to \pi p$, the
situation is different.  Here it is the $\gamma^*_L$ that dominates in
the limit of large $Q$, so that the curves for $\gamma^*_L$ and
$\gamma^*_T$ have to intersect at some value of $Q$.  We do not know
where this is for pion production, and whether it is before or after
the amplitude for $\gamma^*_L$ starts falling again.  In the case of
the $\rho$, where we have extensive data (and many more helicity
transitions), the transitions $\gamma^*_T\, p \to
\rho_T^{\phantom{*}}\, p$ and $\gamma^*_L\, p \to
\rho_L^{\phantom{*}}\, p$ become equal around $Q^2=2 \mbox{~GeV}^2$.

Summarizing this section, there is an intricate relation between
scaling properties and helicity structure in hard-scattering
processes.  It is a consequence of quite general principles. How well
the corresponding helicity selection rules are satisfied at finite
$Q^2$ is governed by the interplay between the hard scale in the
process and the scale of the transverse distribution of partons in a
hadron. It also depends on how important configurations are with
additional gluons in the hadron wave functions.

%==============================================================
\section{Deeply virtual Compton scattering}

\subsection{Why?}

Let me now focus on the exclusive process where theory is most
advanced, and where data from both high and low energies have recently
come in \cite{lob01,bia01,air01,ste01}.  Although this reaction has a
lower cross section than for instance vector meson production, and
although its analysis is more involved due to the competing
Bethe-Heitler mechanism, it presents important advantages.  The first
one is precisely due to the Bethe-Heitler process, since it allows one
to study the real and the imaginary parts of the scattering amplitude
separately through interference \cite{die97,fra97}.

Secondly, to leading and first subleading power in $1/Q$ only the
known pointlike component of the real photon is needed, not the
hadronic part with its unknown wave function.  Let me remark that,
whereas hadronic wave functions do present a source of theoretical
uncertainty in calculating exclusive meson production, this can be
seen as a tradeoff of difficulties.  In hard \emph{inclusive}
processes, calculations are done at the parton level, and in many
cases one must describe the transition from partons to the final-state
hadrons by models with their uncertainties.  In \emph{exclusive}
processes the calculation is already at the level of the final-state
hadrons, and the physics of hadronization is precisely contained in
the meson wave function and similar quantities.

Thirdly, the hard subprocess in Compton scattering involves less
external parton lines than in meson production, and as a consequence
also less internal ones.  As a consequence, the hard external momentum
of the process becomes less ``diluted'' in the hard scattering, and
our experience with exclusive processes suggests that the scale $Q^2$
where factorization works will be lower for Compton scattering than
for meson production.

\subsection{Theory approaches for small $x_B$}

Let me now give a brief survey of the theoretical approaches that have
been applied to DVCS so far, and begin with the model of Frankfurt et
al.\ \cite{fra97}.  Although sometimes referred to as a ``leading
order QCD calculation'', it is \emph{not} the analog of what is
commonly called a LO QCD analysis of $F_2$, i.e., the calculation of
the quark handbag diagram of Fig.~\ref{fig:compton}a with an ansatz
for the quark distribution in the proton.  Instead, it is a rather
simple way of relating the DVCS amplitude with the measured inclusive
structure function $F_2$, which one may represent in three steps:
\begin{displaymath} 
F_2 \propto \mbox{Im}{\cal A}(\gamma^*p\to \gamma^*p) 
\:\stackrel{1\;}{\to}\:
\mbox{Im}{\cal A}(\gamma^*p\to \gamma p)_{t=0}
\:\stackrel{2\;}{\to}\:
{\cal A}(\gamma^*p\to \gamma p)_{t=0}
\:\stackrel{3\;}{\to}\:
{\cal A}(\gamma^*p\to \gamma p)  .
\end{displaymath}
Step 1 is the most involved one.  At low scale $Q^2=Q_0^2$, the ratio
$R$ of $\mbox{Im}{\cal A}(\gamma^*p\to \gamma^*p)$ and $\mbox{Im}{\cal
A}(\gamma^*p\to \gamma p)_{t=0}$ was estimated to be around $0.5$
using an ansatz based on the aligned jet model.  The \emph{variation}
of this ratio with $Q^2$ was then studied using the formalism of
parton distributions and their evolution with the hard scale; this is
where LO QCD entered the argument.  Here the generalized parton
distributions were assumed to equal the usual ones at the low
factorization scale $Q_0^2$.  The authors found little change of $R$
with $Q^2$ and $x_B$.  In step 2, the ratio $\eta$ of $\mbox{Re}{\cal
A}(\gamma^*p\to \gamma p)_{t=0}$ and $\mbox{Im}{\cal A}(\gamma^*p\to
\gamma p)_{t=0}$ was taken from the derivative of the imaginary part
with respect to $\log(1/x_B)$.  This is a widely used approximation,
also in meson production, based on analyticity and the quantum numbers
of two-gluon exchange. It assumes an effective power behavior of the
amplitude in $x_B$; analogous prescriptions for more general
dependence on $x_B$ also exist.  In step 3, an exponential behavior
$e^{B t/2}$ of the amplitude on $t$ was assumed, with a $W^2$ and
$Q^2$ dependent slope $B$ estimated from measurements of other
diffractive processes.

Each of the above steps reflects a rather fundamental aspect in the
physics of DVCS. Step 1 is the transition from a forward to a
nonforward process; in the context of factorization it means going
from a forward to a nonforward parton distribution with its two
independent longitudinal variables.  Step 2 concerns the energy
behavior at small $x_B$ and key approximations made in the high-energy
regime.  As discussed in section~\ref{sec:pictures}, step 3 involves
the question of how longitudinal and transverse dynamics interrelate.
Since the Compton amplitude can in principle be extracted from data,
all three steps can be studied experimentally in DVCS.

There have recently been two applications of the dipole picture to
DVCS. The one by Donnachie and Dosch \cite{don00a} uses a
semiclassical approach to model the dipole cross section at large
dipole size $r$, and two-gluon exchange for small dipoles.  Recently,
McDermott et al.\ \cite{mcder01a} compared the predictions of two
rather different dipole models for this process. It is interesting to
note that both give rather similar results for $\mbox{Im}{\cal
A}(\gamma^*p\to \gamma p)$ in the kinematics of the HERA measurements,
but show more pronounced differences in the real part.  Similar
findings have been made previously by Frankfurt et al.\ \cite{fra98}.

\subsection{Generalized parton distributions}

In contrast to the previous approaches, the leading-twist description
with generalized parton distributions (GPDs) is amenable to both high
and intermediate energies, and thus provides a link between the HERA
Collider and the HERMES measurements.  Let me sketch how currently
predictions are obtained. This proceeds in several steps:
\begin{equation}
q(x;Q_0^2) 
\:\stackrel{1\;}{\to}\:
q(x,\xi,t;Q_0^2)
\:\stackrel{2\;}{\to}\:
q(x,\xi,t;Q^2)
\:\stackrel{3\;}{\to}\:
{\cal A}(\gamma^* p\to \gamma p) .
\end{equation}
Here $q(x,Q^2)$ denotes the conventional quark density at scale $Q^2$,
and $q(\xi,t;Q^2)$ the generalized quark distribution. The procedure
is the same for gluons.  To make predictions we need $q(\xi,t;Q_0^2)$
at some scale $Q_0^2$, and current strategies use the available
parameterizations of $q(x,Q_0^2)$ as a model input.  Let me stress
that what we can currently do in step 1 is to take \emph{model}
prescriptions, mostly based on symmetry considerations. A widely used
strategy is an ansatz based on ``double distributions'' \cite{mus99},
which was also employed in the recent study \cite{fre01}.  The
$t$-dependence is presently modeled by the simple but restricted
factorizing ansatz mentioned in section~\ref{sec:pictures}.

Step 2 is performed using the appropriate evolution equations for
GPDs, whose kernels are known to NLO \cite{bel00a}. Notice that, even
if one takes a given parameterization of the conventional parton
distributions and the same model prescription how to ``convert'' them
into a GPD, the result depends on the scale where this is done.  In
other words, for the model prescriptions we have, translating $q(x)$
into $q(x,\xi,t)$ at $Q_0$ and then evolving up to $Q_1$ does not give
the same as first evolving $q(x)$ from $Q_0$ to $Q_1$ and then
translating it into $q(x,\xi,t)$.  There are arguments \cite{fra97a}
that evolution of the GPDs tends to ``wash out'' the $\xi$ dependence
of the distribution at the starting scale, so that after evolution
over a long interval in $Q$ the $\xi$ dependence is essentially
generated by evolution itself.  This seems to work rather well in the
region $x>\xi$ of the GPDs, but not for $x<\xi$, where evolution works
very slowly, so that it may be most relevant at small $\xi$ and for
observables not sensitive to the region $x<\xi$.

Step 3 above uses the hard scattering kernels, which for DVCS are also
known to NLO \cite{man97}.  The NLO expressions involving a quark box
and generalized gluon distributions are however only known for
massless quarks. For $Q^2$ comparable to $m_c^2$ this leads to an
uncertainty in the results, since there it is neither a good
approximation to omit charm in the quark box, nor to treat it as
massless.  As charm comes with a charge factor $4/9$ in the amplitude
compared with $6/9$ for the sum of light quarks, this uncertainty is
not negligible, and a calculation of the massive quark box for DVCS
would be welcome.

Notice that the imaginary part of the DVCS amplitude is only sensitive
to the point $x=\xi$ at LO, and only to the region $x\ge \xi$ in the
radiative corrections.  The real part probes the entire $x$-region of
the GPDs.  It can involve important cancellations between the regions
$x<\xi$ and $x>\xi$, as is illustrated by the principal value integral
$\int dx\, (x-\xi)^{-1} q(x,\xi,t)$ appearing in the LO expression.

In summary, the theory of Compton scattering belongs to the
best-developed branches in the physics of exclusive processes.  The
phenomenological possibilities of DVCS allow one to study several
important questions rather directly in the data.  It also has a
special status since its theoretical complexity lies in between the
totally inclusive DIS structure functions and the much richer (and
more complicated) processes of meson production.

%==============================================================
\section*{Acknowledgements}

I am indebted to many colleagues for discussions, in particular to
J.~Bartels, W.~Buchm\"uller, J.~Crittenden, K.~Golec-Biernat,
A.~H.~Mueller, R.~Peschanski, and M.~Strikman.  Thanks go to
J.~Bartels for valuable comments on the manuscript.

%==============================================================

\end{document}